\documentclass[review,1p]{elsarticle}
\usepackage{url}
\usepackage[english]{babel}
\usepackage{graphicx} 
\usepackage{multirow} 
\usepackage{array} 
\usepackage[figuresright]{rotating}
\usepackage{float}
\usepackage{tabularx}
\usepackage[table]{xcolor}
\usepackage{amsmath}
\usepackage{graphicx}
\usepackage[colorlinks=true, allcolors=blue]{hyperref}

\makeatletter
\def\ps@pprintTitle{%
 \let\@oddhead\@empty
 \let\@evenhead\@empty
 \let\@oddfoot\@empty
 \let\@evenfoot\@empty
}
\makeatother

\begin{document}
\begin{frontmatter}
\title{Central Bank Digital Currencies: A Survey} 
\author[1]{Qifeng Tang}
\author[1]{Yain-Whar Si\corref{cor1}}

\cortext[cor1]{{Email address: mc35339@um.edu.mo (Qifeng Tang), fstasp@um.edu.mo (Yain-Whar Si, corresponding author).}}

\affiliation[1]{organization={Department of Computer and Information Science, Faculty of Science and Technology, University of Macau},
            city={Macau}, 
            country={China}}

\begin{abstract}
With the advancement of digital payment technologies, central banks worldwide have increasingly begun to explore the implementation of Central Bank Digital Currencies (CBDCs). This paper presents a comprehensive review of the latest developments in CBDC system design and implementation. By analyzing 135 research papers published between 2018 and 2025, the study provides an in-depth examination of CBDC design taxonomy and ecosystem frameworks. Grounded in the CBDC Design Pyramid, the paper refines and expands key architectural elements by thoroughly investigating innovations in ledger technologies, the selection of consensus mechanisms, and challenges associated with offline payments and digital wallet integration. Furthermore, it conceptualizes a CBDC ecosystem. A detailed comparative analysis of 26 existing CBDC systems is conducted across four dimensions: system architecture, ledger technology, access model, and application domain. The findings reveal that the most common configuration consists of a two-tier architecture, distributed ledger technology (DLT), and a token-based access model. However, no dominant trend has emerged regarding application domains. Notably, recent research shows a growing focus on leveraging CBDCs for cross-border payments to resolve inefficiencies and structural delays in current systems. Finally, the paper offers several forward-looking recommendations for future research.
\end{abstract}

%
%



\end{frontmatter}

\section{Introduction}

The 2008 financial crisis eroded public confidence in centralized financial institutions. During this period, private digital currencies, represented by Bitcoin, emerged \cite{han2024revolutionizing, raskin2024private}. The increasing influence of digital currencies has put pressure on governments to safeguard public sovereignty \cite{peneder2022digitization}. This has driven central banks to conduct research on their own digital currencies. Various factors contribute to the development of Central Bank Digital Currencies (CBDCs) in different countries, such as ease of payment, financial inclusion, and the search for stable means of payment \cite{auer2020rise}. At present, numerous nations are actively researching (CBDCs) and deliberating on their design and implementation \cite{darbha2022archetypes, kosse2023making, Riksbank2023_ekronapilot}. The widespread adoption of mobile payments reflects growing consumer demand for secure and efficient digital transaction methods.

There is currently no universally accepted definition of central bank digital currency (CBDC) \cite{allen2020design}. The Bank for International Settlements (BIS) initially referred to CBDC as a general concept rather than a specific term \cite{coere_loh_2018}, and later defined it as a new form of digital currency, denominated in national currency units and directly issued by the central bank \cite{boar_wehrli_2021}. Ward and Rochemont \cite{ward2019understanding} described CBDC as a digital representation of central bank money, distinct from traditional reserve or settlement accounts. Bjerg \cite{bjerg2017designing} emphasized its characteristics as "electronic," "universally accessible," and "central bank issued." For depositors, CBDC may function as a substitute for interest-bearing commercial bank deposits \cite{carapella2020central}. A more concise definition frames CBDC as a digital currency issued by national or regional central banks \cite{10.1145/3510487.3510498}.

The primary objectives of CBDC issuance include promoting financial inclusion, enhancing monetary policy effectiveness, improving financial stability, and increasing the security and efficiency of both domestic and cross-border payments \cite{lee2021survey, bc2020cbdcfpcr}. Enabled by advances in computer technology \cite{10.1145/3510487.3510498}, CBDCs act not only as policy instruments but also as catalysts for the transformation toward more secure, inclusive, and efficient payment systems \cite{rachmad2025cbdc}.
Unlike commercial bank deposits, CBDCs are issued by central banks, denominated in national currency units, and backed by sovereign credit. Some implementations utilize distributed ledger technology (DLT) to enhance transaction speed. CBDCs can be stored in digital wallets and do not necessitate a traditional banking relationship. As noted by Bech et al. \cite{bech_garratt_2017}, decentralization enables CBDCs to facilitate peer-to-peer transactions in a manner similar to cash.

CBDCs and electronic money exhibit functional similarities, both provide convenient means of payment and help reduce transaction costs \cite{fabris2019cashless}. However, CBDCs further demonstrate the ability to maintain monetary value \cite{pillai2024central}. Compared to cryptocurrencies\cite{vardhini2024cryptocurrency}, the monetary policy of a CBDC is determined by the central bank, while the currency exchange system may operate independently of that policy. Regardless of the underlying technology, CBDCs are centralized in terms of regulation. In contrast, the monetary policy of cryptocurrencies is embedded in the underlying blockchain protocol, which simultaneously governs both protocol rules and transaction execution. Cryptocurrencies are decentralized, emphasize privacy autonomy, and are characterized by high price volatility.

A 2023 survey conducted by the Bank for International Settlements (BIS) revealed that approximately 94\% of responding central banks are actively engaged in CBDC research \cite{di2024embracing}. Among them, 81\% are developing proofs of concept, and around 33\% have launched pilot projects. Several countries — e.g., Jamaica, the Bahamas, and Nigeria — have already issued CBDCs. In major economies, research has also advanced considerably. For example, China is conducting a large-scale pilot of its Digital Currency Electronic Payment (DC/EP), known as the e-CNY, and is participating in cross-border CBDC initiatives like Project mBridge, in collaboration with Thailand, the United Arab Emirates, and others.

While most central banks have initiated CBDC-related work, the degree of progress varies significantly across jurisdictions. As state-backed digital currencies, CBDCs must incorporate robust safeguards for infrastructure and user privacy. These concerns necessitate design considerations not only prior to issuance but also through continuous technological improvements to mitigate evolving vulnerabilities.


\subsection{Data Collection}
\begin{figure}
    \centering
	\includegraphics[height=10cm]{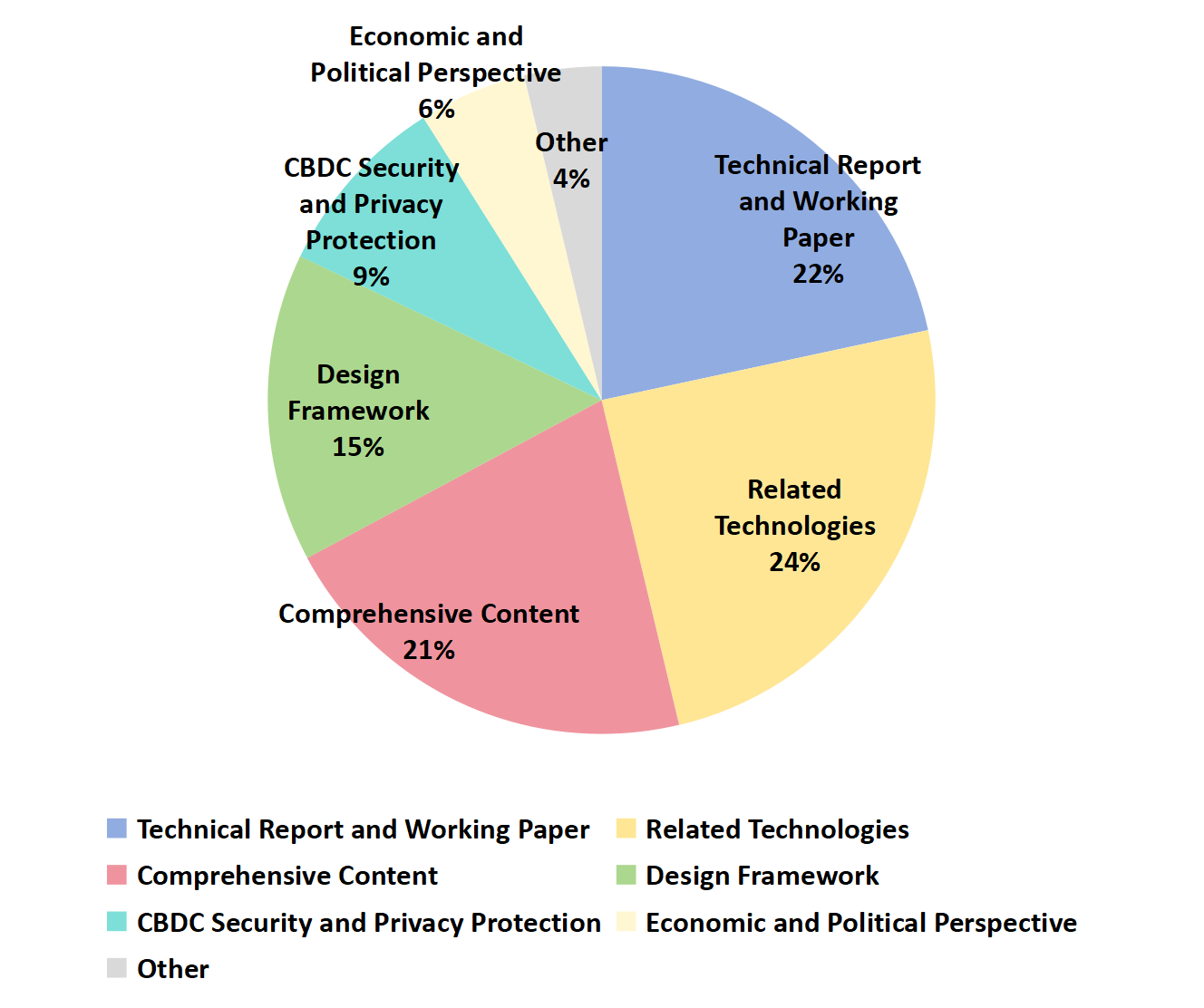} 
	\caption{Categories of the papers cited in this review.}
	\label{fig:chart1}
\end{figure}

In this paper, an extensive literature review on CBDCs was conducted, and the initial search yielded 346 articles. To ensure the quality of citations, we screened the literature based on citation counts. The final selection consisted of 135 articles which are chosen based on relevance, quality, timeliness, and completeness of data. Categories of the papers cited in this review is depicted in Figure \ref{fig:chart1}.

\begin{itemize}
\item As shown in Figure \ref{fig:chart1}, 22\% of the 135 cited articles are research reports and working papers from authorities such as the Bank for International Settlements (BIS) and central banks. These organizations remain the primary authoritative sources of research on CBDCs, and their articles describe experiments, pilots, implementation experiences, and problem summaries of CBDC systems by authorities in various regions and countries, including central banks and technical agencies involved in CBDC development. 

\item 24\% of the articles focus on related technologies, detailing technologies that can be applied to or have already been integrated into CBDC systems, such as Distributed Ledger Technology (DLT) and blockchain.

\item 15\% of the articles are dedicated to CBDC system design offering an in-depth exploration of the design taxonomy. System design remains a central focus of research, as there is currently no standardized offering an in-depth exploration of the design taxonomy for CBDC system design. Much of the research continues to explore different design possibilities. Moreover, research on related technologies often focuses on their systematic integration to effectively fulfill the objectives of CBDC design. Few articles address the optimization of existing systems.

\item 9\% of the articles focus on the security and privacy aspects of CBDCs. These articles primarily discuss how to secure CBDC systems and protect user privacy, emphasizing the critical importance of these issues.

\item 6\% of the articles examine the economic and political perspectives of CBDCs. These articles explore the political, economic, and social impacts of CBDC systems.

\item 21\% of the articles adopt a multi-perspective approach, addressing CBDC systems from the aspects of technology, design, privacy, security, and economic policy. These articles explore the intersection of multiple factors and their impact on the development and implementation of CBDCs.

\item The remaining 4\% cover other topics, primarily research methodologies, relevant techniques, and citations from authoritative literature.
\end{itemize}

\begin{figure}[htbp]
	\includegraphics[width=1\textwidth]{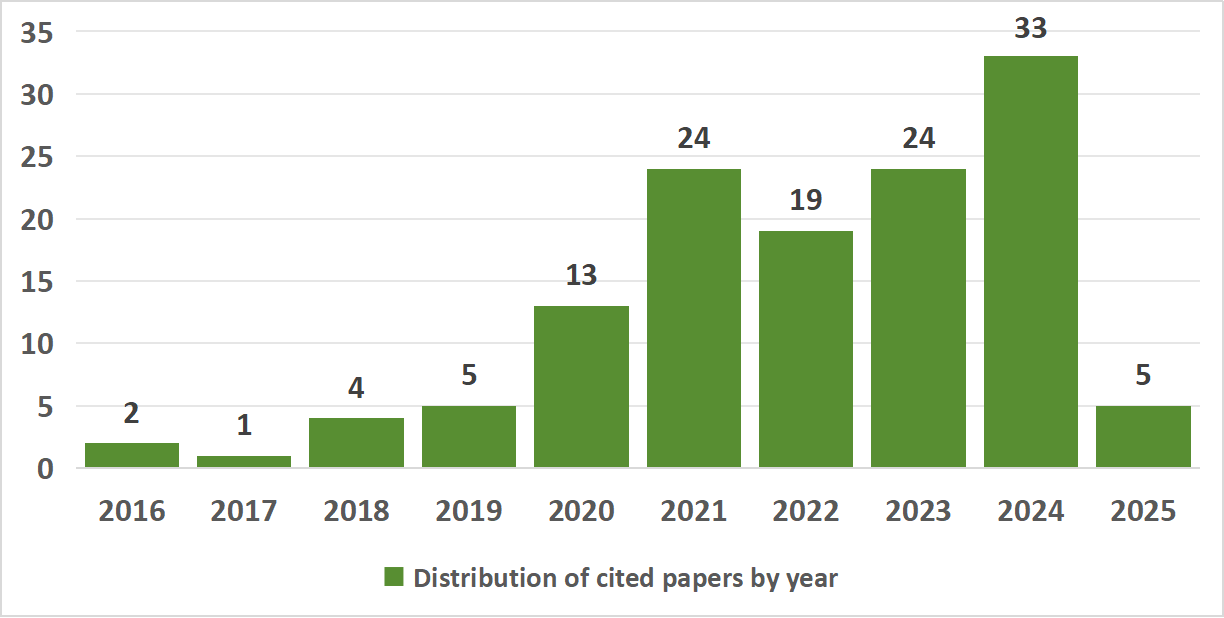} 
	\caption{Distribution of cited papers by year.}
	\label{fig:chart2}
\end{figure}

The time distribution of the cited papers is shown in Figure \ref{fig:chart2}, spanning from 2016 to 2025. The majority of these articles were published between 2020 and 2024, with only 12 articles published before 2019. Among all the articles, four are official dynamic web pages, which are continuously updated over time and thus were not included in the temporal statistics for this analysis.

We have also summarized the 38 most recent publications on CBDCs for the years 2024-2025, as shown in Table \ref{table: paper distribution}. It can be seen that the majority of the articles originate from technical reports and working papers issued by central banks and international institutions. In addition to these, a substantial number of studies focus on enabling technologies and system architecture, particularly design taxonomies and implementation frameworks. By contrast, relatively few works address the broader economic and regulatory implications of CBDC deployment. The design and implementation of CBDCs involve multifaceted challenges, including technical feasibility, privacy protection, financial inclusion, and system integration. Experimental projects across various countries and regions have provided valuable practical experience, highlighting key issues and potential technical solutions that will be crucial in the future design and adoption of CBDCs.

\begin{table}[!ht]
    \footnotesize
    \begin{tabularx}{1\textwidth}{|p{6.61cm}|p{6cm}|}
    \hline
        Type & Papers from 2024 to 2025 \\ \hline
        Technical Report and Working Paper & \cite{noll2024observations},  \cite{Jordan2024HelvetiaIII}, \cite{Faykiss2024}, \cite{Bank_of_Japan_2024}, \cite{CBE2024}, \cite{RBA2024}, \cite{MFP2024}, \cite{BOT_2024}, \cite{bakx2024digital} \\ \hline
        Related Technologies & \cite{han2024revolutionizing}, \cite{guo2024dlt}, \cite{cryptoeprint:2024/1206} , \cite{dumbre2024blockchain}, \cite{bespalova2024crypto}, \cite{michalopoulos2025privacy}, \cite{atangana2025securing}, \cite{goudjil2025application}, \cite{cryptoeprint:2024/1746}\\ \hline
        Comprehensive Content & \cite{di2024embracing}, \cite{vardhini2024cryptocurrency}, \cite{iwanczuk2024potential}, \cite{RePEc:bis:bisbps:151}, \cite{SanzBayón2025}, \cite{central_bank_of_the_bahamas_2024}, \cite{CORBET2024106226}, \cite{freiman2024cbdc}, \cite{li2025current} \\ \hline
        Design Taxonomy & \cite{pillai2024central}, \cite{tsareva2023cbdc}, \cite{nyffenegger2024proposal}, \cite{mishra2024simple}, \cite{bowler2024non}, \cite{lamberty2024hybcbdc}, \cite{genc2024literature}  \\ \hline
        Economic Implications and Policy Considerations & \cite{raskin2024private}, \cite{MBAKOB2024101005}, \cite{bjt2023cbdcfi}  \\ \hline
    \end{tabularx}
    \caption{Papers from 2024 to 2025.}
    \label{table: paper distribution}
\end{table}

\subsection{Contributions}
This article surveyed the latest CBDC-related articles and made the following contributions:
\begin{itemize}
    \item \textit{Extended CBDC System Design Taxonomy}. Building upon the existing CBDC design pyramid, this study refines and extends the design taxonomy by incorporating updated insights from 135 academic and official publications. The enhanced taxonomy provodes a clear perspective on CBDC system design, helping subsequent researchers to quickly grasp the architecture of CBDC systems.
    
    \item  \textit{Conceptualization of the CBDC System Ecosystem}. Beyond technical classifications, this paper proposes a comprehensive CBDC system ecosystem that integrates system architecture, ledger technology, access model, and application scenarios. This ecosystem-oriented view supports a more holistic understanding of CBDC design and deployment.
    
    \item \textit{Summary of 26 CBDC Projects Design Projects.} This study reviews and organizes the design features of 26 existing CBDC projects along four key dimensions: system architecture, ledger technology, access model, and application areas. These design choices are presented in a table, revealing that the most frequently adopted combination is a two-tier architecture with distributed ledger technology and a token-based access model.
    \item \textit{Forward-Looking Research Recommendations.} Finally, the paper outlines several forward-looking directions for future research
\end{itemize}

\subsection{Organization}
This paper is structured as follows: Section 1 provides an introduction to the paper and an overview of CBDC systems. Section 2 describes specific aspects of CBDC system design and outlines the ecosystem of a CBDC system. Section 3 discusses the global status of CBDC design and the current challenges faced. Section 4 presents a brief discussion on the security and privacy protection of CBDC systems as well as their economic and political implications. Finally, Section 5 concludes the paper.

\section{CBDC System Design Taxonomy} 
\begin{figure} 
    \centering
	\includegraphics[height=7cm]{{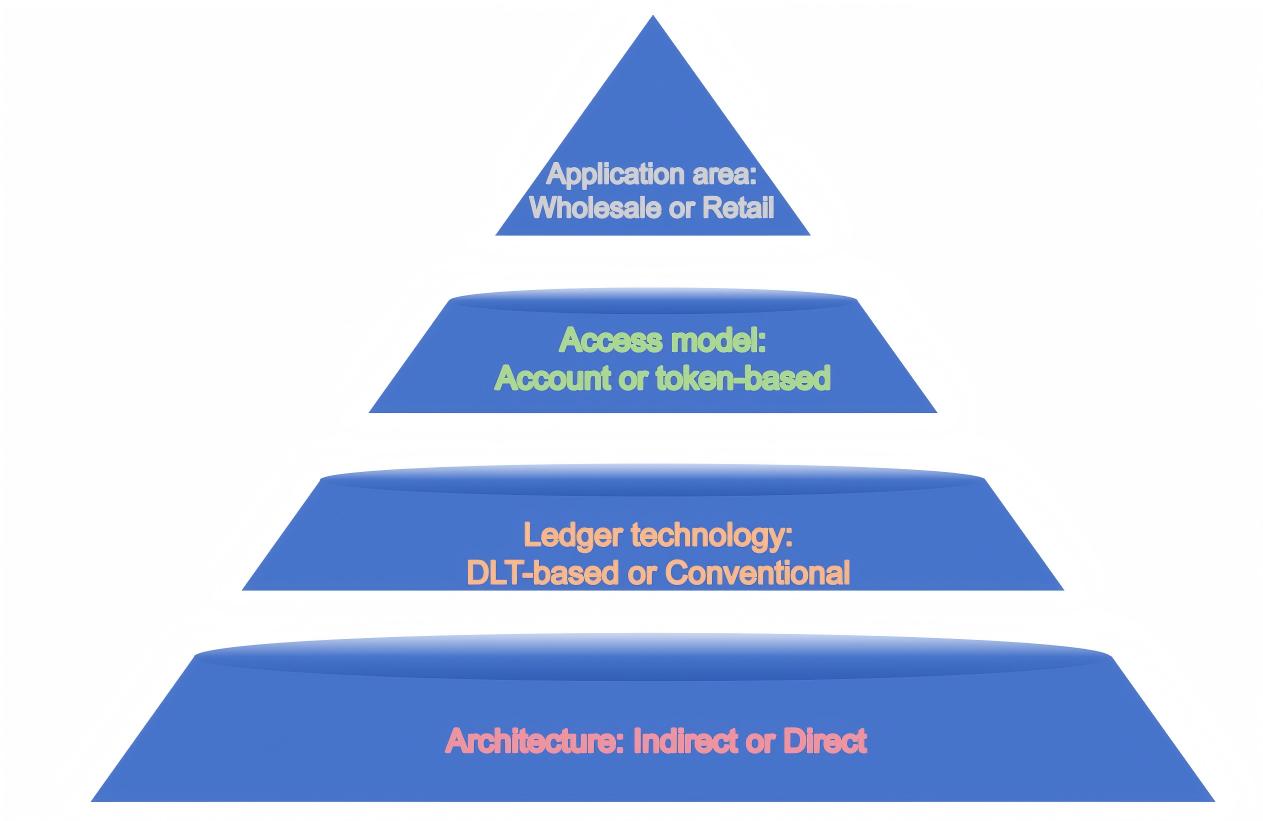} } 
	\caption{The CBDC taxonomy pyramid, adopted from \cite{auer2020technology}. } 
	\label{fig:pyramid} 
\end{figure}

The fundamental taxonomy of CBDC systems is intrinsically linked to user needs, with key design dimensions interacting dynamically \cite{auer2020technology}. As illustrated in Figure \ref{fig:pyramid}, CBDC systems are structured around four core design dimensions. At its foundation is the system architecture, which includes three types: direct, indirect, and hybrid. In a direct model, the central bank manages all transactions and balances. The indirect model relies on intermediaries to hold user accounts, with the central bank overseeing them. The hybrid model blends both approaches—offering flexibility and complexity—allowing customization based on the roles of the central bank and intermediaries. This architectural choice influences the selection of ledger technology, which can be either distributed (DLT) or centralized (CLT). Regardless of the approach, the CBDC system must remain secure and resilient.


The ledger infrastructure determines data update mechanisms while directly impacting privacy and security. It also determines how users access the system—either through identity verification (account-based) or by presenting an cryptographic token (token-based). The access model further impacts the complexity of cross-border settlements, which must be built on the wholesale layer of CBDC. In terms of application area, CBDCs are divided into retail (for the general public) and wholesale (for high-value financial transactions).

\begin{sidewaysfigure}
    \centering
    \includegraphics[width=\textwidth]{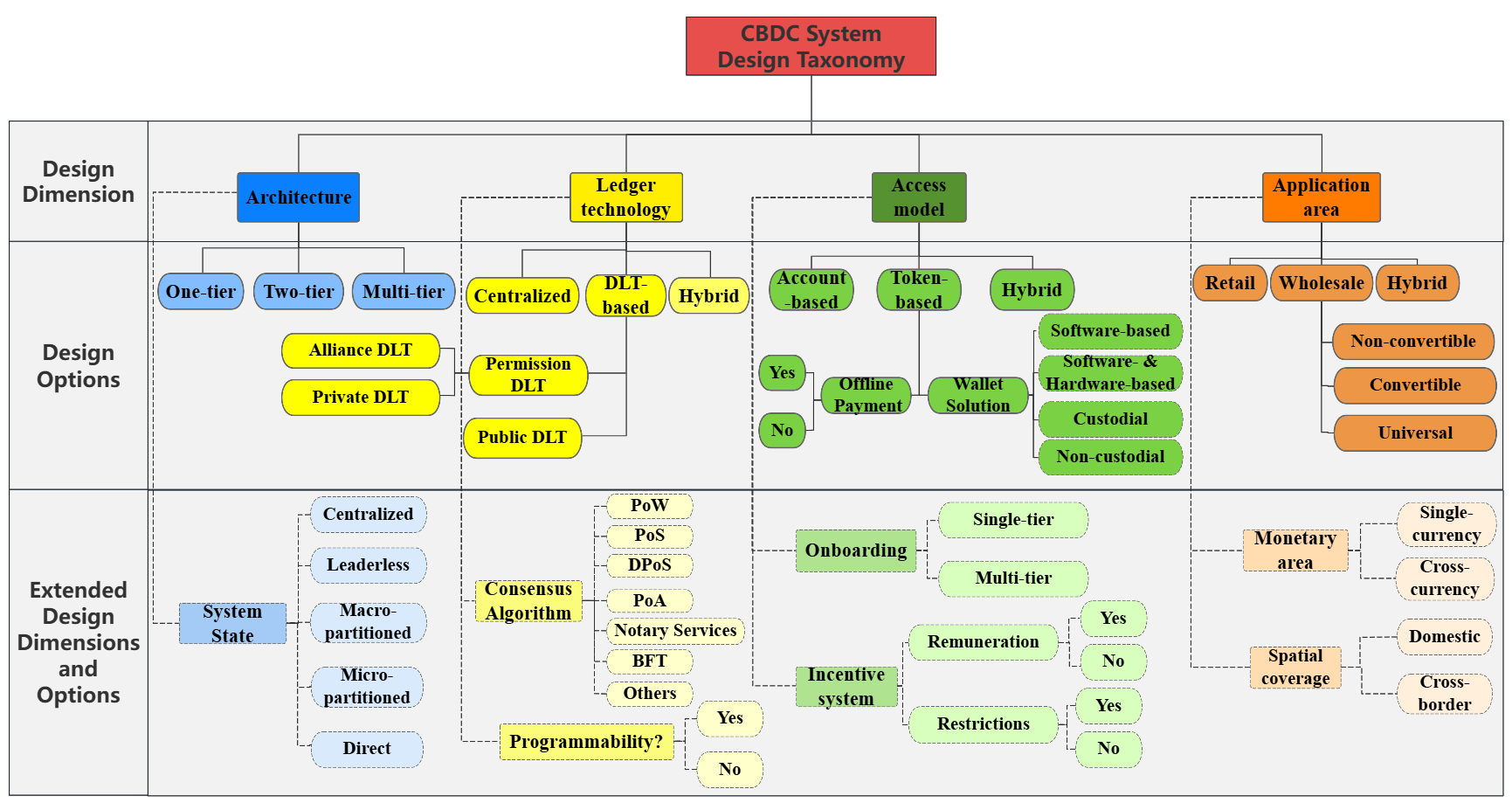}
    \caption{Taxonomy of CBDC, summarized from \cite{auer2020technology, stamm2023taxonomy, darbha2022archetypes, guo2024dlt, zhang2022blockchain, tsareva2023cbdc, kochergin2020central, bowler2024non}.}
    \label{fig:taxonomy}
\end{sidewaysfigure}

In designing a CBDC, additional factors beyond the four core dimensions in Figure \ref{fig:pyramid} must also be considered. As shown in Figure \ref{fig:taxonomy}, the taxonomy pyramid diagram refines Figure 4 further refines the taxonomy by elaborating on each core dimension—such as distinguishing between permissioned and public DLT under Ledger Technology—and introducing extended dimensions including consensus algorithms, programmability, onboarding, and spatial coverage. This taxonomy offers a flexible and comprehensive guide for tailoring CBDC systems to diverse technical, regulatory, and policy needs. The following sections will explore each aspect in depth.

\subsection{System Architecture} 
The choice of CBDC architecture influences the system's functions, processing efficiency, security, and its impact on financial systems \cite{tsareva2023cbdc, TERCEROLUCAS2023101188, MBAKOB2024101005}. It also determines the complexity of implementing a CBDC system and its various system properties. In general, the architecture is divided into one-tier and two-tier systems, although some have proposed architectures with three or more tiers \cite{albychev2023development, 9795279}, which however can be regarded as a further refinement of the 2-tier system. 


Mirroring traditional monetary systems—which separate interbank and non-bank circuits via commercial banks—CBDC architectures are commonly categorized into one-tier or two-tier models, as shown in Figure \ref{fig:taxonomy}. In a one-tier model, the central bank interacts directly with end users, offering simplicity and real-time control but imposing a heavier operational load on the central bank. In contrast, the two-tier model delegates user-facing functions to financial institutions, easing the central bank's burden while maintaining oversight. Indirect and hybrid architectures fall under this category, often improving scalability and user experience through institutional collaboration.

A common feature is the further decentralization of some transaction processing responsibilities to relevant CBDC agencies, The additional tier can also serve as a privacy feature \cite{9795279}. Through dynamic virtual addresses or other privacy-protecting mechanisms, relevant organizations can conceal the true identity of their customers, thereby protecting trade secrets. Only the regulator and the institution are aware of the mapping between the virtual address and the real customer identity. For example, assigning certain CBDC functions to single or multiple operating agencies \cite{albychev2023development}. A single operator enables uniform transaction processing but poses higher risks of privacy breaches and data centralization. In contrast, multiple operators mitigate data monopolies through decentralization, though they increase system complexity and regulatory oversight demands.\\

 Furthermore, system design can also be analyzed through the distribution and maintenance of system state across participants. A complementary dimension of CBDC system classification concerns the distribution and maintenance of system state across participants. As illustrated in Figure \ref{fig:taxonomy}. One such classification includes five models\cite{darbha2022archetypes}: Centralized, Leaderless, Macro-partitioned, Micro-partitioned, and Direct.

In the Centralized model, a single entity maintains the state, allowing for easier regulatory oversight, legal data access, and efficient support for online transactions and system scalability, albeit with reduced user privacy. The  Leaderless system distributes control through consensus among multiple entities, providing similar benefits to centralized models with improved resilience via data replication. However, it compromises privacy and scalability as data volume increases. The Macro-partitioned model distribute transaction processing across multiple entities, each managing a partition that may include shared states. They support online payments but provide limited privacy. The Micro-partitioned systems divide the state into smaller segments, some controlled by users, enhancing privacy but increasing complexity in system expansion and online operations. As these models lack global state replication and require central coordination. In the direct model, transactions are overseen solely by the participants without third-party involvement, resembling cash. This approach enhances privacy, data security, and offline capability, but limits online functionality and regulatory oversight. This model is generally suitable only for low-value transactions. 

Each model reflects trade-offs among privacy, scalability, regulatory compliance, and operational complexity.  The choice of system operating status is not only influenced by the system architecture but also closely related to ledger technology.
 
\subsection{Ledger Technology} 
The interaction among CBDC participants determines whether infrastructure and access rights are centralized or decentralized, directly influencing the design and operation of the core ledger, as discussed in subsequent sections. The choice of ledger technology influences system performance metrics such as scalability, throughput, latency, and fault tolerance. The core ledger also records issuance data and transaction flows \cite{be2023dptwp}. As shown in Figure \ref{fig:taxonomy}, CBDCs are typically categorized into three types based on their ledger infrastructure: Centralized Ledger Technology-based CBDC (CLT-Based CBDC), Distributed Ledger Technology-based CBDC (DLT-Based CBDC), and hybrid CBDC \cite{wang2021international}. CLT-based CBDC relies on a centralized database for transaction recording and validation, aligning closely with conventional account-based banking systems. In contrast, DLT-based CBDC uses blockchain or other distributed ledgers to enable secure, transparent transactions within a decentralized network. The following subsections examine the three major categories of ledger technology in CBDC systems: centralized, distributed, and hybrid models.

\subsubsection{Centralized Ledger Technology(CLT)} 
A Centralized Ledger (CL) refers to a system where a single trusted entity manages and validates all transactions, typically using traditional techniques for centralized recording and authorization \cite{app13010499}. CLT-based CBDC systems require a trusted institution—typically the central bank—to oversee and maintain the ledger. Due to their strong alignment with monetary governance, CLs are widely favored by countries designing CBDCs. These systems rely on centralized databases, often using relational models such as MySQL, Oracle, or SQL Server, which are standard in financial and enterprise applications.


\paragraph{Limitations of CBDC based on CLT} When utilizing CLT, the storage and management of data are centralized at a single central node, with other nodes only able to retrieve or send data to the center. This architectural pattern gives rise to several key drawbacks:

\begin{itemize}
\item \textit{Excessive Power of the Central Node} \cite{deng2022review}. The central node holds significant control, with limited external oversight, potentially compromising transparency, user rights, and system security.

\item \textit{Single Point of Failure (SPF) Risks}. The system's reliance on one central node creates a vulnerability—its failure may halt the entire system. Although primary-backup architectures, redundancy backups, or disaster recovery mechanisms can mitigate this, the SPF risk remains inherent.

\item \textit{Limited Privacy} \cite{app13010499}. Centralized storage grants default access to all transaction records, heightening privacy risks. While access control, data minimization, or encryption can help, the risk is not eliminated.

\item \textit{Lack of Rigorous Verifiability} \cite{yang2022ubiquitous}. Centralized control limits users' ability to independently verify data integrity, especially in cases of tampering.

\end{itemize}

These issues collectively increase the risk of data leakage. While Data Leakage Prevention (DLP) systems are commonly employed, relying on DLP alone may be insufficient. It is important to consider combining DLP with other tools, such as zero-knowledge proofs (ZKP) and differential privacy techniques \cite{mukati2022role, goudjil2025application}.


\paragraph{CLT Innovation for CBDC Systems}

Proposed in 2020\cite{10.14778/3415478.3415540}, LedgerDB is a centralized ledger (CL) database offering blockchain-like features such as tamper resistance, non-repudiation, and enhanced auditing, while maintaining data consistency and security through centralized management. It presents a viable alternative to permissioned blockchains.

Its architecture includes three core components: Ledger Master, which manages metadata and coordinates cluster-level events. Ledger Proxy, which receives, pre-processes, and routes client requests to the appropriate servers, and can directly access the storage layer to handle transaction payloads. Ledger Server, which completes request processing and stores data in the underlying storage layer.

Like traditional centralized databases, LedgerDB enables fast transaction processing through a single control point, resulting in low latency under stable network conditions. Its throughput surpasses that of blockchain systems. LedgerDB also offers blockchain-like features, including timestamping, tamper-evidence, non-repudiation, and auditing capabilities for tracking user behavior and operational traces. It supports verifiable data deletion, such as purging obsolete records to save space and hiding records to meet regulatory demands, enhancing its real-world applicability. However, reliance on third-party timestamping for external audits poses potential risks. Additionally, while auditability of deleted data is supported, it introduces technical complexity, higher maintenance costs, and potential challenges in ensuring data consistency and audit accuracy.

\subsubsection{Distributed Ledger Technology(DLT)} 
Recent CBDC research increasingly favors permissioned distributed ledger platforms that ensure settlement finality—meaning transactions are irrevocable—and incorporate robust data privacy protections \cite{edwin2020design}. In contrast to CLT, DLT is a ledger shared across a network of nodes, where the consistent state of transactions is stored and replicated at each node \cite{guo2024dlt}. DLT offers superior security features \cite{sethaput2023blockchain, rohmalia2023designing}, mainly by reducing data integrity risks, availability risks, and privacy risks related to system access \cite{lee2021survey}. 

Blockchain technology is a desirable method for implementing DLT. In blockchains, all data is stored in discrete blocks, and these blocks are linked in a chain. Each new child block is linked to the previous parent block through a hash value. During this process, all previously generated blocks are verified \cite{guo2024dlt}. Individual nodes in the blockchain maintain a complete record of valid blocks. Since some blockchains do not have supervisory institutions, a consensus mechanism is necessary to ensure the system functions properly. Consensus mechanisms are also used to ensure the consistency and validity of block data in blockchains supervised by a central node. The four core components of blockchain are: block, chain, network, and smart contract.

Blockchain-based DLTs offer synchronized and immutable records across nodes \cite{kumar2021permission}. These features facilitate auditing and regulation, making blockchain an ideal choice for CBDC. In addition, the adoption of blockchain-based CBDC contributes to cost reduction, improved payment efficiency \cite{zhang2022blockchain}, and enhanced cross-border payments \cite{jung2021blockchain}. Blockchain implementation relies heavily on cryptography and consensus algorithms \cite{esposito2021blockchain}. The ISO/IEC 11179 metadata registry eas proposed to be used in blockchain-based CBDC systems to ensure proper CBDC transactions and interoperability \cite{jung2021blockchain}. 

\paragraph{DLTs used in CBDC} Typically, there are public, consortium, and private blockchains \cite{zhang2022blockchain}. Among them, consortium chains and private blockchains are collectively referred to as permissioned blockchains, i.e., blockchains that require permission from the relevant organization to be accessed. Anyone can participate in public blockchains without permission, where UTXO (Unspent Transaction Output) is used to prevent double spending \cite{zhang2021hybrid}. Specifically, each transaction output can only be used once, thus ensuring the uniqueness and non-tampering of the transaction.

Several studies have reflected on the adoption of permissionless block-chains for CBDC \cite{Berger2023, guo2024dlt, nyffenegger2024proposal}. While this would reduce development costs, transaction costs are higher, and central banks are subject to the network rules of the blockchain. To address these issues, less reliance on the underlying blockchain rules is required \cite{nyffenegger2024proposal}. A consortium blockchain consists of multiple organizations federated together as nodes of the blockchain, and authorization is required to access the blockchain. In contrast, a private blockchain is controlled by a single organization that manages the ledger maintenance.

Permissioned blockchains are more aligned with the regulatory requirements needed by CBDC \cite{kumar2021permission}. Moreover, a CBDC based on a Permissioned Blockchain Network (PBN) can reduce SPF risks and achieve zero downtime through redundancy and high availability design, enabling commercial banks to interact with each other without going through the existing Real-Time Gross Settlement System (RTGS). Currently available platforms for permissioned blockchains include, but are not limited to, Corda, Hyperledger Fabric, Hyperledger Iroha, Hyperledger Besu, Elements, Interledger, and Bitt \cite{sethaput2023blockchain}. Each platform has its own characteristics, and the right platform should be chosen based on the requirements for CBDC. 

As for the choice of CBDC structure, a characterization of the different blockchains is presented in Table \ref{table:DLT structure}. Features such as cross-border transaction speed, interoperability, transparency, trust, security, and decentralization are present to a high degree in public DLT, a medium degree in consortium DLT, and a low degree in private DLT. In contrast, features such as domestic transaction speed, throughput, and scalability are high in private DLT. Consortium DLT and private DLT require permission. 

\begin{table} 
	\includegraphics[width=\textwidth]{ 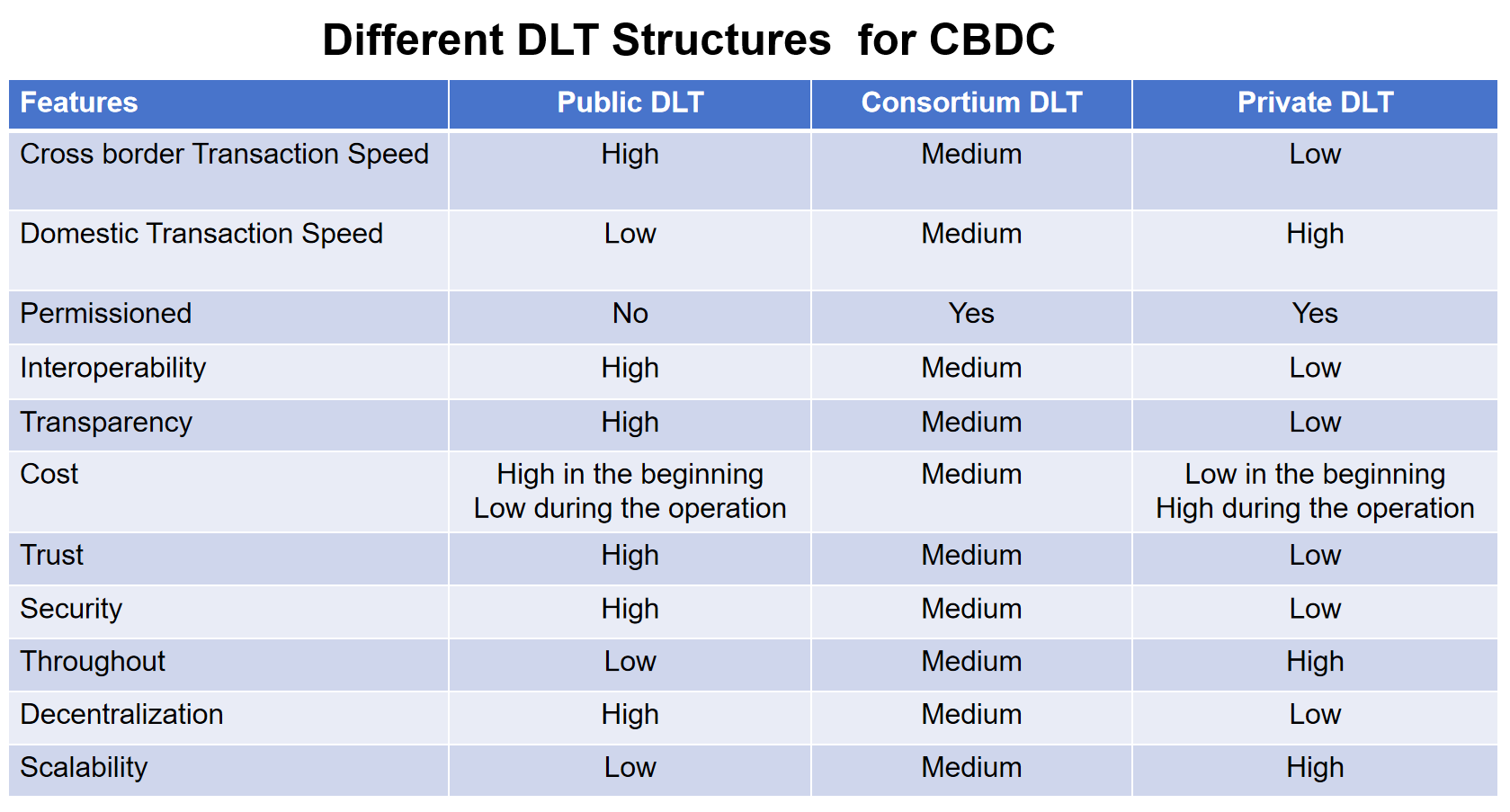} 
	\caption{DLTs used in CBDC, adopted from \cite{guo2024dlt, zhang2022blockchain}. } 
	\label{table:DLT structure} 
\end{table} 

In terms of cost, public DLTs are more expensive to be developed in the initial phase, mainly due to the need for complex consensus mechanisms and higher computational resources. However, in the maintenance phase, the cost may decrease as the network size stabilizes. Private DLTs are easier to deploy in early stages but tend to incur higher maintenance costs as they scale. Consortium DLT is typically at an average level. In addition to the features in Table \ref{table:DLT structure}, it is essential to consider the following issues when choosing a DLT for CBDC \cite{guo2024dlt}: Byzantine Fault Tolerance (BFT), Immediate Block Finality (IBF), and Smart Contract Support (SCS), to ensure that the CBDC's technical foundation is secure and reliable. 

On the other hand, some scholars \cite{tsai2018multi, tsai2016system} categorize blockchains used in CBDCs and cryptocurrencies into two types: ABC (Account Blockchain) and TBC (Trading Blockchain). As their names imply, ABC stores and transmits account information, while TBC manages transaction execution and related data. Separating accounts from transactions across multiple nodes enhances scalability and manageability, and reduces transmission paths, thereby lowering the risk of information leakage. However, this approach also increases system complexity and requires more detailed design.


\paragraph{Consensus algorithms, smart contracts, and interoperability}Consensus algorithms are essential in ensuring consistency across distributed nodes. The choice of consensus algorithm is crucial for CBDC, as it directly affects the system’s performance, scalability, security, and governance \cite{tsareva2023cbdc}. The core features to consider for CBDC's consensus algorithm are security, finality, and efficiency, in addition to trade-offs between transaction throughput and decentralization \cite{tsareva2023cbdc}. The decentralization of CBDC is within the system and is further divided into political, architectural and logical decentralization. Typically, CBDCs are politically and logically centralized and architecturally decentralized (Transaction throughput refers to the number of transactions that are capable of being processed per unit of time).

Comparison of consensus algorithms adopted in CBDCs is given in Table \ref{table:consensus algorithms}. The consensus algorithms include Proof of Work (PoW), Proof of Stake (PoS), Proof of Authority (PoA), Delegated Proof of Stake (DPoS), and Notary Service (Validating Notary Service and Non-validating Notary Service) \cite{tsareva2023cbdc}.

\begin{table} 
	\includegraphics[width=\textwidth]{ 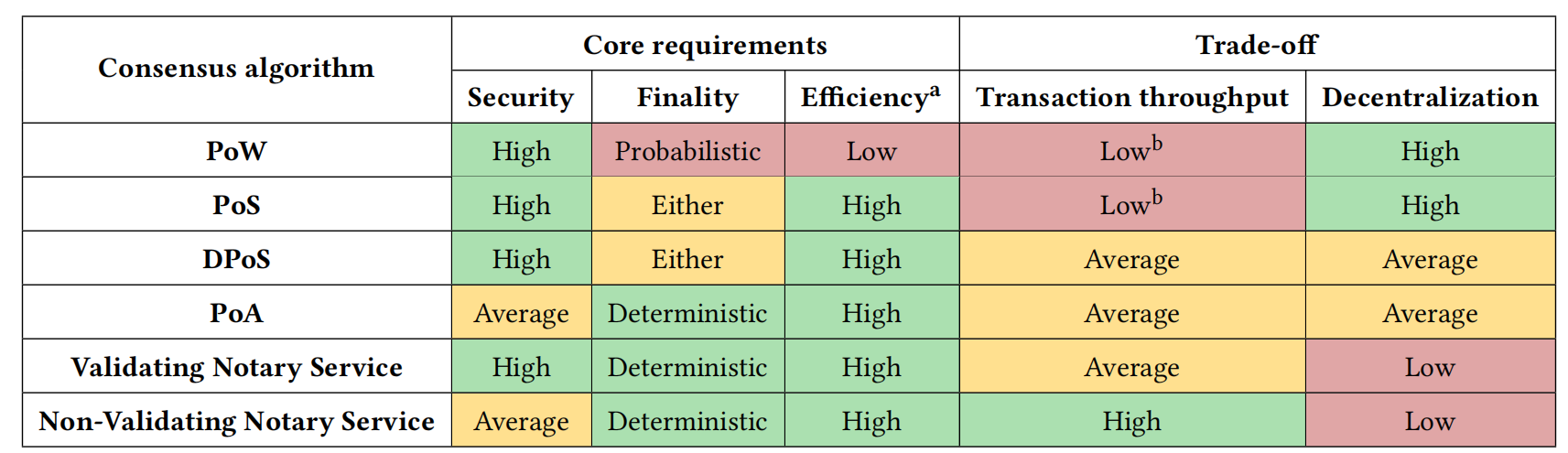}
	\caption{Comparison of consensus algorithms used in CBDCs, adopted from \cite{tsareva2023cbdc}. } 
	\label{table:consensus algorithms} 
\end{table}  
 
PoW secures the ledger by requiring nodes to solve computational puzzles, ensuring integrity but incurring high energy costs and low throughput. Its security stems from the economic infeasibility of controlling over 51\% of network power, though it introduces risks from hidden attackers and limited scalability.

PoS systems, validators are selected either randomly based on pledged cryptocurrency or automatically upon meeting a pledge threshold. This ensures a fair and decentralized allocation of validation rights while preserving network security and integrity. Depending on the method, transaction finality can be deterministic or random. PoS is energy-efficient and offers high throughput but may lean toward centralization, as validator power correlates with token holdings or stakes.
    
    
DPoS achieves high efficiency by allowing token holders to elect a small set of validators to generate and verify blocks, though at the cost of reduced decentralization. Its security relies on economic disincentives for malicious behavior. As an extension of PoS, DPoS offers similar finality but higher throughput due to fewer participating nodes.
    
    
 PoA relies on a group of pre-screened or certified nodes to validate transactions, offering energy efficiency and fast confirmation but with limited decentralization. Security is maintained by these trusted nodes, which is generally acceptable in CBDC systems.
    
    
Notary Service ensures transaction uniqueness via a central bank-operated notary node. It is efficient and offers fast confirmation but is the least decentralized. There are two types: non-validating, which only records transactions, and validating, which also verifies them. The authenticated type enhances security in untrusted networks but lowers throughput, while the unauthenticated type assumes trusted nodes and achieves higher throughput.
    

Given CBDC's need for efficiency, finality, and energy conservation, PoA, DPoS, and notary-based mechanisms are more suitable than PoW or PoS, aligning better with regulatory and operational requirements.

Another widely adopted consensus mechanism in CBDC system is Byzantine Fault Tolerance (BFT), which enables distributed networks to achieve agreement even in the presence of faulty or malicious nodes. It is resilient to various faults, including malicious behavior, and can tolerate up to one-third of nodes failing or acting dishonestly. A well-known variant, Practical Byzantine Fault Tolerance (PBFT), has been implemented in real-world CBDC trials such as South Africa's Project Khokha \cite{ProjectKhokha2}. PBFT emphasizes high security and strong consistency, making it well-suited for stable, permissioned environments with known participants.


Smart contracts are self-executing programs deployed on the blockchain that automate transactions based on predefined conditions, reducing the need for manual intervention. The programmability of smart contracts allows processes that originally required manual operations to be automatically executed, and enables them to be modified according to different business needs. In the CBDC system, smart contracts are used in transactions, settlements, order processing, and more. Their traceability can reduce the risk of fraud and default \cite{franko2022towards}. Additionally, compared to users directly deploying smart contracts, achieving the liquidity of CBDC in smart contracts through integrated solutions, such as bridging, may be more reliable and secure \cite{kocsis2021research}. However, the adoption of DLT technology has introduced certain obstacles to financial crime investigations and conflicts with some data privacy regulations, such as the General Data Protection Regulation.


\paragraph{Data security} For Distributed Ledger Technology (DLT)-based CBDCs, data security is a critical concern. Whether the CBDC runs on a permissioned or permissionless blockchain, the primary security threats are spoofing, tampering, and information leakage \cite{hans2023blockchain}. Alternative solutions to address these issues include Secure Multiparty Computation, Zero Knowledge Proof, Ring Signature, and Homomorphic Encryption \cite{lee2021survey}. In addition to these approaches, data encryption plays a particularly essential role in solving these problems. 
To enhance data security beyond traditional encryption, commitment schemes such as the Pedersen Commitment Scheme\cite{wang2022model} can be applied. It protects ledger privacy, verifies transaction amounts, and ensures integrity by allowing participants to commit to a value without revealing it, while preventing subsequent alterations. With the advancement of quantum computing, traditional encryption algorithms face increasing vulnerability. To mitigate this threat, post-quantum cryptographic algorithms—designed to resist quantum attacks—should be considered to ensure long-term system security \cite{cryptoeprint:2024/1206}.


\subsubsection{Hybrid Models} 

While blockchain and DLT are frequently discussed in the context of CBDCs, full decentralization is often deemed incompatible with central control over monetary policy, user privacy, and data governance. Given the uncertainty around blockchain's suitability for CBDC, hybrid models have gained attention from researchers and policymakers.



Scholars have proposed a digital currency model combining account-based fund management with electronic cash transaction processing, realized in the Platypus system \cite{10.1145/3548606.3560617}. In Platypus, each participant holds a central bank-signed account commitment containing a serial number and balance. During transactions, both parties reveal their account serial numbers and use zero-knowledge proofs to verify authenticity, ensuring that the total balance remains unchanged.This allows for completely anonymous transactions \cite{10.1145/3548606.3560617}. This approach not only protects transaction privacy but also simplifies regulatory implementation. Leveraging traditional account databases improves transaction speed and lowers client-side requirements. However, the system remains vulnerable to network-level anonymization and de-anonymization attacks, as adversaries may still re-identify individuals by correlating external data.

Similarly, Danarto Tri Sasongko and Setiadi Yazid proposed a CBDC design methodology in 2020 that integrated both DLT and non-DLT systems \cite{10.1145/3427423.3427447}. The system combines DLT and non-DLT technologies, allowing the public to use wholesale DLT-based CBDC. It enhances interbank settlement efficiency, strengthens monetary control, and lowers user costs, while mitigating bank disintermediation. The two-tier architecture involves central bank-commercial bank transactions in the first tier and CBDC distribution to the public by commercial banks in the second. These banks act as DLT nodes, develop user-facing applications, and maintain token accounts for unallocated funds. The DLT chain features immutable, ordered transaction data across interconnected areas, which cannot be efficiently analyzed. Transaction data is stored and synchronized via relational or NoSQL databases \cite{muzammal2019renovating}.\\

\subsection{Access Model} 

While ledger technology discusses how this data is recorded and managed, the access model  focuses on who can access this data, to what extent, and under what conditions. As shown in Figure \ref{fig:taxonomy}, user access patterns fall into three categories: account-based, token-based, and hybrid. Account-based CBDCs require identity and account verification, whereas token-based models operate independently of identity systems. Instead, they verify the authenticity of the digital tokens to conduct the transaction \cite{ozili2023central}. 

\subsubsection{Account-based Model} 
Even though decentralized cryptocurrencies are growing rapidly, many people still prefer account-based access over token-based CBDCs when using CBDC \cite{abramova2022can}. Account-based CBDC, also known as register-based CBDC \cite{mishra2024simple}, relies on payment technology enabling instant, low-cost settlements. The system operator authenticates users and grants permission to modify accounts or CBDC only after verifying appropriate access rights. Account-based models are considered highly viable due to their balance between data privacy and payment system security \cite{BIS2021AER}. CBDC is transferred between accounts using a centralized database similar to traditional banks \cite{ozili2023central}, with user fund information and transaction records kept in the account. It is typically capable of handling high transaction volumes and is suitable for large-scale use \cite{urbinati2021digital}.

Moreover, account-based CBDC can limit CBDC holdings, even if they are controlled and regulated by the central bank. Identity traceability facilitates (AML) enforcement supervision, and the risk of CBDC in cross-border circulation may be effectively reduced \cite{BIS2021AER}. However, the issue of recognition of identity information across different countries still needs to be addressed. Overall, account-based models offer strong regulatory benefits but face international interoperability challenges.

\subsubsection{Token-based Model} 
Another model is the token-based system, which works similarly to cryptocurrencies like Bitcoin \cite{bofinger2020cbdc}. In token-based CBDC systems, tokens are usually transferred without going through a centralized hub, with the initiator of the transaction transferring the tokens directly to the receiver. The token contains all the information needed for the recipient to verify the legitimacy of the transaction. The recipient of the token is authorized to validate the token to ensure that it is secure and usable \cite{digitaldollarproject2020}. 

To maintain consistency and prevent double spending, token models often restrict token divisibility or implement tamper-evident mechanisms \cite{urbinati2021digital}. For instance, a \$10 token cannot be split into two \$5 tokens, ensuring transaction consistency and preventing double spending. Moreover, as token-based CBDC doesn't rely on a centralized ledger and uses digital signatures for verification, signatures can typically be validated locally, enabling offline transactions.Tokens generally offer more privacy, but they usually require a distributed ledger, which complicates the payment system \cite{ozili2023central}. 


Moreover, the security of token-based CBDC mainly depends on private keys, which can be inconvenient for individuals to manage \cite{kahn2020security}. Loss or theft of a private key can result in the loss of digital assets. As private keys are complex character strings, managing them requires technical skills such as backup or hardware wallet usage. A practical solution is needed, with e-wallets being a common option.

User wallets are broadly categorized as custodial and non-custodial \cite{bowler2024non}. In custodial wallets, CBDC is stored off-device and managed by a third party. This model is more user-friendly and suitable for novices due to its simplicity and available support, though users must still manage accounts and passwords securely and remain aware of potential risks \cite{voskobojnikov2021u}. In contrast, non-custodial wallets give users full control over assets and private keys, reducing reliance on third parties but requiring greater technical proficiency to ensure key security. CBDC wallets are typically software-based, but offline transactions often require integrated hardware and software solutions, as shown in Figure \ref{fig:taxonomy}.

Offline payments are essential for financial inclusion and system resilience \cite{cryptoeprint:2024/1746}, yet they pose significant privacy and compliance challenges. A promising approach combines hierarchical system architecture with auditable anonymity to balance these concerns \cite{michalopoulos2025privacy}. To strengthen trust, central banks should avoid holding sole decryption authority, instead employing secure multi-party computation (MPC) or threshold encryption. Additionally, delayed synchronization and post-transaction compliance reporting mechanisms can reduce the risk of misuse in offline environments.

Privacy-preserving technologies—such as anonymous credentials, zero-knowledge proofs, and verifiable encryption—can conceal sensitive identity and transaction data while enabling detection of double spending through auditability. Robust private key management is equally important. Biometric authentication (such as fingerprint or facial recognition) enhances usability, while social recovery mechanisms support secure key retrieval through trusted contacts or recovery phrases, mitigating asset loss due to key mismanagement.



\subsubsection{Hybrid Access Model} 

A CBDC system can incorporate a token-based design even with authentication in place. To enhance flexibility, it may combine account-based and token-based access models, each with distinct advantages and limitations. A hybrid approach helps offset their respective drawbacks—for instance, account-based transactions suit large or business-to-business payments, while token-based ones are better for micropayments and retail use.

One such implementation of the hybrid model is HybCBDC \cite{lamberty2024hybcbdc}. It integrates account-based and UTXO-based subsystems to balance confidentiality and transparency through a hybrid access model. The account-based layer handles wholesale and retail transactions, akin to real-time gross settlement systems, while the UTXO-based layer supports retail-level private payments, with UTXOs functioning as tokens. These subsystems are linked via the Interchain Communication Protocol (IBC), enabling CBDC transfers within and across layers. Financial institutions oversee currency conversion between the two subsystems.


In the context of CBDC, programmability (Figure \ref{fig:taxonomy}) refers to the ability to define the behavior of a currency in digital form through a computer program \cite{lee2021programmable}. Currency programmability enables embedded functions within the payment chain and can be implemented via traditional accounts or digital tokens—through APIs in the former and smart contracts in the latter.

Additionally, user access to the CBDC model is inevitably tied to the onboarding process, as shown in Figure \ref{fig:taxonomy}, which involves user registration and authentication. There are single-tier and multi-tier structure types \cite{stamm2023taxonomy}. To implement risk hierarchy management, the central bank may choose a multi-tiered user verification system. This system provides users with different levels of validation options to enable risk hierarchy management \cite{BankOfGhana2023, pbc2021prdec}. Lower levels of authentication generally mean that users are more restricted in their use of CBDC, such as transaction limits or wallet balance restrictions. As authentication levels increase, users gain higher privileges and more lenient restrictions. In contrast, single-tier onboarding mechanisms provide the same authentication message for all users \cite{stamm2023taxonomy}. Such systems are typically simpler but limited in scope. These choices relate to Know Your Customer (KYC), a process where institutions verify user identities to mitigate risks like fraud and money laundering, applying transaction privileges based on verification levels.


The adoption of distributed ledger technologies such as blockchain can simplify the KYC process \cite{dumbre2024blockchain}. Traditional KYC processes require repeated authentication across institutions, resulting in redundancy, inefficiency, and higher data leakage risk. In a DLT-based CBDC system, once verified, a customer's KYC information can be shared across nodes, eliminating the need for resubmission and reducing duplicate data collection. This streamlined approach often relies on smart contracts for implementation.

Additionally, many digital currencies have incentives, such as Bitcoin's mining rewards and Ether's transaction fees. The CBDC design can consider incentives both in terms of remuneration and restrictions \cite{stamm2023taxonomy}. One can consider whether there should be interest generation or transaction restrictions.

\subsection{Application Area}
The design of the application Area is essentially to determine the object-orientation of the CBDC. Retail CBDC is a public-facing digital currency used for daily transactions and is intended to complement existing banknotes and cash \cite{luu2023implications, genc2024literature}. In contrast, wholesale CBDC is used for large-value payments and settlements between financial institutions \cite{luu2023implications}.

\subsubsection{Retail CBDC} 
The purpose and roles of retail CBDC vary across different countries. In developed countries, the issuance of retail CBDC aims to improve payment efficiency, security, and resilience \cite{kosse2022gaining}. Meanwhile, the motives for issuing retail CBDCs in developing countries focus on promoting financial inclusion, improving payment efficiency, and ensuring effective implementation of monetary policy \cite{bjt2023cbdcfi, iwanczuk2024potential, kosse2022gaining, auer2020rise}.

Typically, retail CBDCs are legal tender, serve as a means of payment, function as a cash supplement, operate on a secure and adaptable payment infrastructure, ensure interoperability and full usability, and offer users a competitive ecosystem with instant, secure, and cost-effective services \cite{morales2021implementing}. However, some scholars argue that retail CBDC may not only serve as a supplement to cash but also as a substitute for or operate in parallel with cash \cite{kochergin2020central}. Nevertheless, retail CBDCs share some characteristics with cash, with token-based retail CBDCs ensuring anonymity. It should be noted that, unlike virtual currencies, the anonymity of retail CBDCs is subject to limitations to restrict illicit behavior, but there is currently no standard or quantitative guideline for an acceptable level of anonymity \cite{kochergin2020central}.

Retail CBDCs offer diverse and technically feasible design options. They may be directly managed by the central bank or distributed via financial institutions. Ledger choices include traditional centralized systems or DLTs like private blockchains. Currency can be stored in bank accounts or digital token wallets. As retail CBDCs serve the public, funds should be accessible 24/7 \cite{kochergin2020central}.


In addition to retail CBDCs, fast payment systems (FPS) offer real-time or near-real-time retail transactions, typically using commercial bank or electronic currencies. Table \ref{table 4} compares the two, highlighting their distinct features and complementary roles in policy support. Both enable instant transactions and multiple channels, but retail CBDCs emphasize financial inclusion and stability, while FPS is more effective for cross-border payments.


\begin{table} 
	\includegraphics[width=\textwidth]{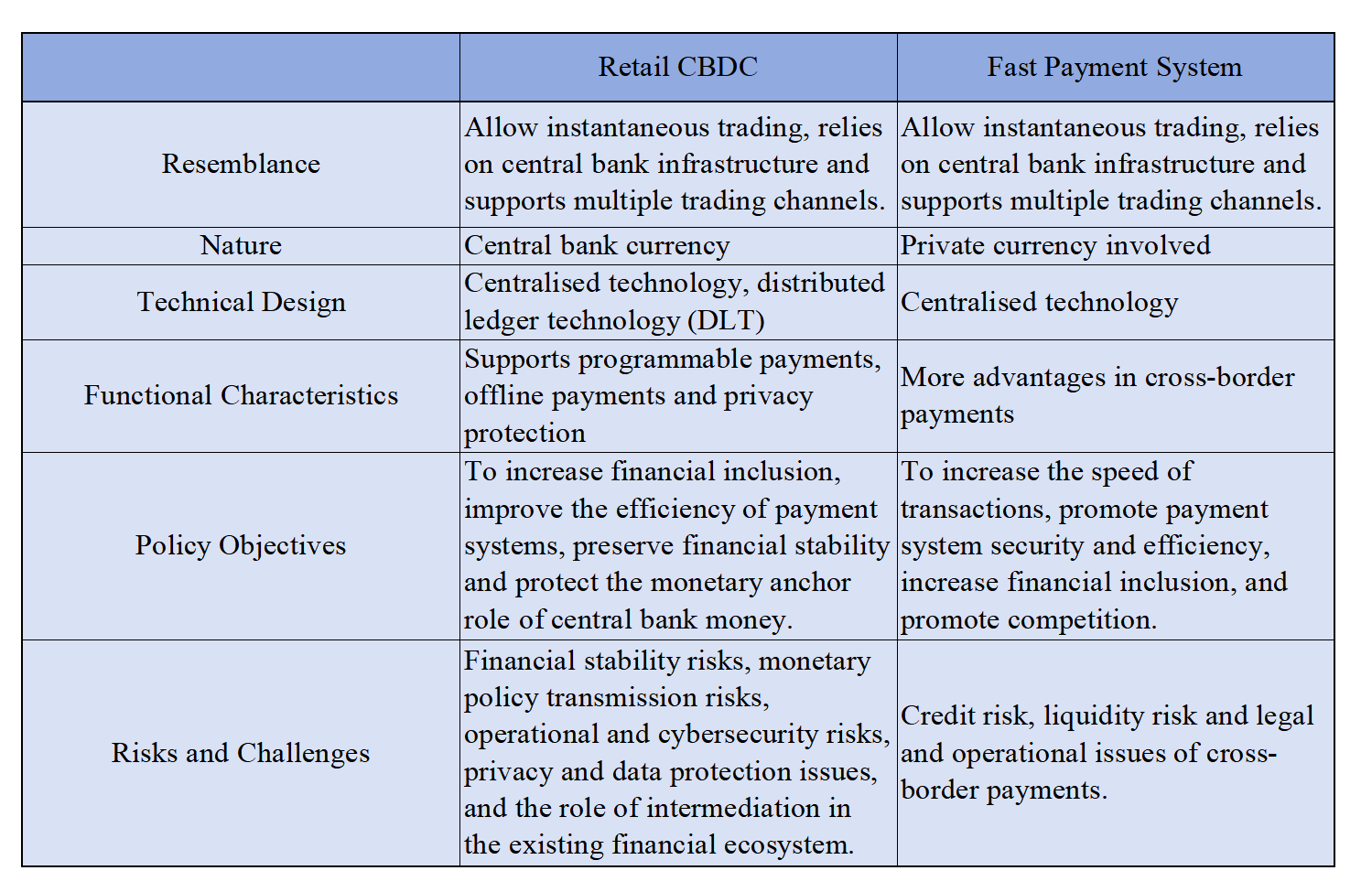}  
	\caption{Comparison of Retail CBDC and FPD, adopted from \cite{RePEc:bis:bisbps:151}. } 
	\label{table 4} 
\end{table} 

\subsubsection{Wholesale CBDC} 
Wholesale CBDC are used by banks or financial institutions for mutual payments and are not accessible to the public \cite{fsb2018analysis}. Unlike retail CBDC, wholesale CBDCs often involve a hybrid model for access and storage, such as tokens stored under a bank account \cite{kochergin2020central}. Wholesale CBDCs are primarily used for large-value digital currency transactions and cross-border payments. For most countries that have issued or plan to issue wholesale CBDCs, the goal is to improve the efficiency of cross-border payments \cite{kosse2022gaining}. Wholesale CBDCs, which usually use DLT, can enhance the efficiency of cross-border payments by addressing existing problems in cross-border bank payments, such as high costs, slow speeds, and lack of transparency \cite{de2021wholesale}. Due to the large amounts involved, wholesale CBDCs are generally not traded by non-banking institutions. 

Wholesale CBDCs can be categorized into three types \cite{kochergin2020central}: non-convertible, limited to a single jurisdiction; convertible, allowing exchange between two jurisdictions; and universal, designed for multi-country circulation. Their technical implementation is increasingly complex compared to RTGS modifications. Domestic wholesale CBDCs are typically non-convertible, while cross-border models fall under convertible or universal types. Among these, the generic wholesale CBDC model shows the greatest potential for addressing cross-border payment inefficiencies \cite{kochergin2021central}. However, its adoption may be constrained in countries with underdeveloped financial infrastructures due to substantial technological, financial, and operational demands.


Cross-border CBDC systems must account for multi-currency transactions. Differences in technical standards and infrastructure across jurisdictions pose challenges to interoperability, a key factor in reconciling currency and ledger disparities \cite{SanzBayón2025}. Traditional solutions rely on trusted intermediaries, introducing reliance on trust. One alternative is using smart contracts to define cross-book payments between script-enabled ledgers, while non-scripted ledgers require bridge ledgers \cite{wood2016polkadot, zamyatin2019xclaim}. However, this approach confines users to specific environments, adds trust assumptions, and increases SPF risk. Another method introduces a middle layer that abstracts transaction authorization, enabling ledger-independent cross-ledger applications \cite{cryptoeprint:2023/116}. While this simplifies development, such middleware must be carefully designed to prevent non-compliant operations like money laundering.

Earlier, we discussed the spatial coverage and currency areas of CBDCs \cite{stamm2023taxonomy}, referring to the countries where CBDCs are used and the total amount in circulation. The spatial coverage of CBDCs may not align with currency boundaries. For instance, the Eurozone spans multiple countries using a single currency, the Euro. Meanwhile, countries like France and Canada are developing their own CBDCs, suggesting that multiple CBDCs could coexist within a single jurisdiction.


Platforms such as Corda, Quorum, and Hyperledger Fabric are commonly used in wholesale CBDC development \cite{edwin2020design}, offering enhanced transparency, security, and immutability. Distributed ledger technology mitigates SPF risks by replicating transaction records across multiple nodes.


\subsubsection{Hybrid of retail CBDC and wholesale CBDC} 
Both CLT-based CBDC and DLT-based CBDC have their own strengths and drawbacks. The main advantages of retail CBDC may include providing users with a highly liquid, low-risk, and universally available means of payment \cite{kochergin2021central}. The main advantages of wholesale CBDCs include faster, safer, and cheaper cross-border payments \cite{kochergin2021central}. Most CBDCs are designed to minimize risk by choosing a hybrid model of the two. 

Huseyin Oguz Genc and Soichiro Takagi \cite{genc2024literature} propose a hybrid CBDC in which the central bank issues and manages the base currency, while payment processing is handled by financial institutions or payment service providers (PSPs)—essentially a 2-tier model. This design mitigates bank disintermediation risks and ensures the involvement of institutions like commercial banks. The digital euro design program is also based on a combination of the two models \cite{urbinati2021digital}. This model combines a CLT-based platform (TIPS+) with a token-based DLT system (e.g., itCoin). Similarly, China’s e-CNY adopts a hybrid CBDC approach, offering the convenience of retail use while ensuring wholesale-level security \cite{genc2024literature}.

The scope of application of CBDCs is also considered in terms of spatial coverage and monetary area \cite{stamm2023taxonomy}. This involves considering whether CBDCs are used domestically or across borders, and whether a single currency or multi-currency model is used. There are three main types of cross-border CBDC use: one through intermediary media, another through interfaces to achieve interoperability between different systems, and the third using a unified CBDC system.

\subsection{CBDC Ecosystem} 
\begin{figure}
    \centering
    \centerline{\includegraphics[height=10cm]{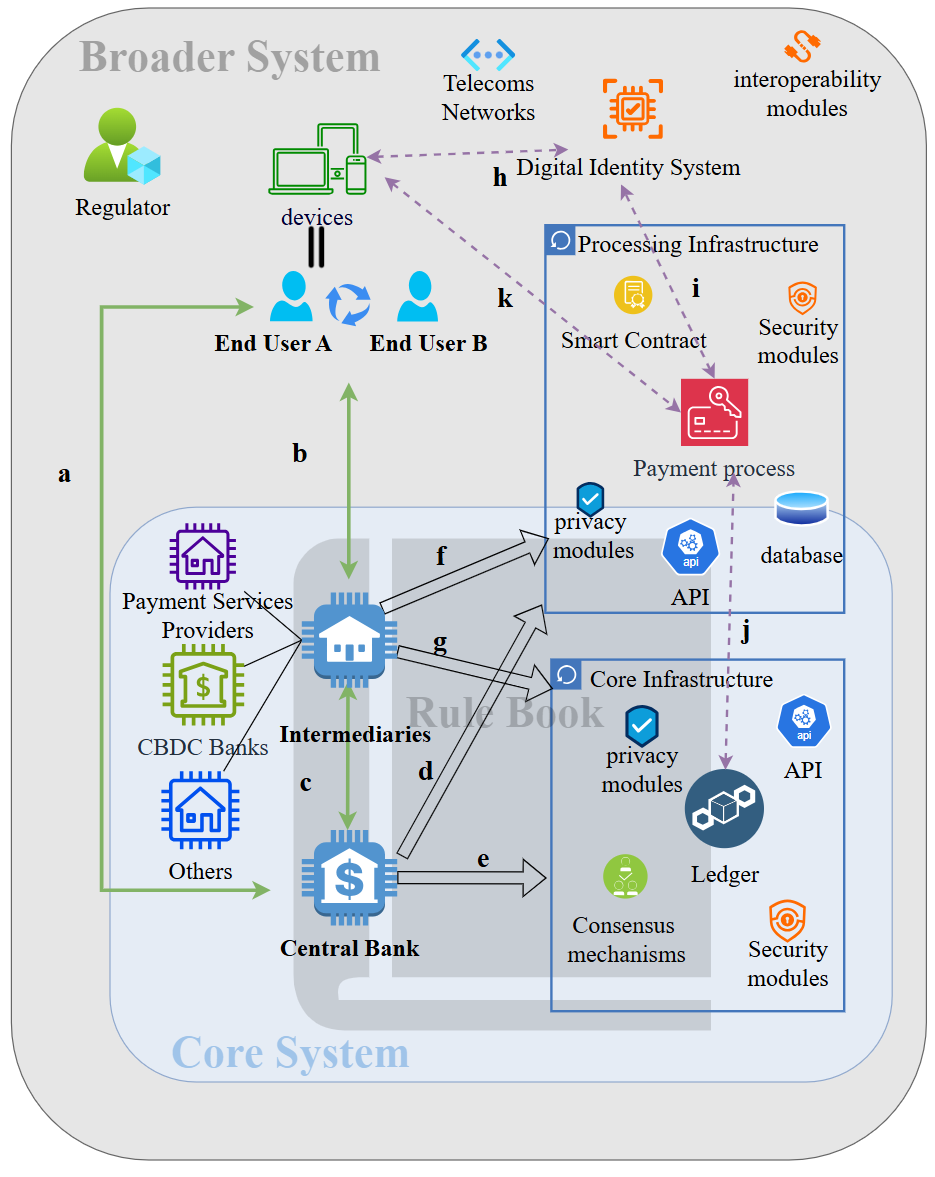} } 
    \caption{CBDC ecosystem. } 
    \label{fig 5} 
    \end{figure}

Building on prior discussions and the Bank of Canada's statement \cite{CBDC2021}, we provide a conceptual overview of the CBDC system architecture using Figure \ref{fig 5}. The figure offers a comprehensive view of the ecosystem, illustrating both the interactions among participating entities and the technical support provided by the underlying ledger. This helps to better understand the operational mechanism of the CBDC system.

As shown in Figure \ref{fig 5}, the CBDC ecosystem comprises a Core System and a Broader Ecosystem \cite{CBDC2021}. The core system includes the Core Rulebook, set by the central bank and binding for all participants, and the Core Infrastructure, which together provide the system's foundation and operations. The broader ecosystem covers additional infrastructure and entities supporting CBDC functionality beyond the core. Figure \ref{fig 5} highlights the key participants in the CBDC ecosystem: end-users, central banks, and intermediaries.

End-users conduct transactions and payments, including individuals, merchants, and financial institutions depending on the application context. The central bank designs, manages, and oversees the system to ensure security, stability, and policy alignment. It also handles issuance and redemption of CBDCs and may assume intermediary functions in some models. Intermediaries serve as key bridges in the CBDC system, responsible for circulation, user registration, payment processing, and customer service. Their roles vary by system architecture and may include banks, payment service providers, and other financial institutions. The ecosystem also involves supporting entities like regulators, though they are beyond the scope of this section. The operation of the CBDC ecosystem relies on the collaboration of rules and infrastructure, which includes, but is not limited to, the following key components:
\begin{itemize}
\item The Rulebook is the legally and technically based framework within which the entire system operates \cite{CBDC2021}. The central bank defines the legal basis, governance framework, risk management, and participant access requirements within the Rulebook.

\item The Core Infrastructure \cite{CBDC2021} consists of the Core Ledger and all components responsible for ensuring that the Core Ledger is operational, transactionally secure, and tamper-proof. This includes, but is not limited to, smart contracts, privacy components, and security features. The core ledger is a key part of the core infrastructure, which also supports CBDC issuance, redemption, and settlement. It enables the central bank to monitor, implement remuneration mechanisms, and exercise centralized control. While the central bank typically manages the core infrastructure, DLT-based systems may involve intermediaries in its operation.


\item The Processing Infrastructure, a key component of the broader ecosystem, handles CBDC payment processing tasks such as message preparation, processing, and reconciliation. It links end-user devices to the core infrastructure, ensuring efficient transmission of transaction data—essential for system stability. While typically operated by intermediaries, the central bank may also be involved in 1-tier or systems with a mix of direct and indirect CBDC.


\end{itemize} 
Beyond the Processing Infrastructure, the broader ecosystem includes user terminals, digital identity systems, and interoperability modules for integration with other CBDC or payment systems. These components are generally managed by intermediaries or third parties. The role interaction pattern in the CBDC ecosystem depends on its architectural design (see Figure \ref{fig 5}). There are three main scenarios:

\begin{itemize} 
    \item the central bank directly interacts with users (a), processes all real-time payments, and records CBDC holdings. It also operates both the underlying and processing infrastructures (d+e). It is a direct CBDC architecture. 
    
    \item intermediaries handle consumer interactions and exchange necessary information with the central bank (b+c). The central bank manages the core infrastructure (e), while intermediaries operate the processing facilities (f) and may assume additional roles defined in the rulebook (f+g). In some cases, the central bank also addresses infrastructure-related functions (f+d). This structure is indirect.

    \item In a hybrid architecture, central banks are directly involved in some of the CBDC transaction processing (a+b+c). The central bank and intermediaries interact with the infrastructure in a similar way. This architecture combines direct and indirect models. 
\end{itemize}

When a user completes a transaction, the interaction path follows h–i–k–j. Specifically, end-user A initiates a transfer to B via a device, triggering identity verification through the digital identity system (h). If authentication fails, a failure notice is returned (h); if successful, payment processing begins (i). A failed transaction returns an error via path k; if successful, a confirmation is sent (k), the data is stored, and relevant information is synchronized to the core ledger (j).


\section{Recent CBDC Systems} 

 
According to the Atlantic Council \cite{atlanticcouncil_cbdctracker}, as of March 2025, three countries have issued CBDCs, and 44 countries or currency unions are piloting or have piloted CBDC projects. Many experiments aim to validate technological feasibility and emphasize payment system stability and security. For example, Project Aber \cite{Aber2020}, based on Hyperledger Fabric, explored cross-border payments using DLT. Beyond payment applications, it also proposed future uses such as serving as an RTGS backup, settling other assets (e.g., bonds), and enhancing interoperability with international central banks.


The Bank of Thailand's Retail CBDC Pilot Project \cite{BOT_2024} highlights CBDC's potential in retail payments, particularly for financial innovation and programmable payments, while revealing challenges such as business model sustainability for non-bank institutions, offline device stability, and dual payment handling. Similarly, Hungary's Digital Student Safe project \cite{Faykiss2024} examines CBDC's role in promoting financial inclusion among students. The Bank of Japan's pilot \cite{Bank_of_Japan_2024} has advanced in technology and functionality testing, demonstrating real-world feasibility but still faces issues related to privacy, system integration, and regulatory compliance.

User experience and market acceptance are critical to CBDC adoption. The Bank of Japan's pilot \cite{Bank_of_Japan_2024} gathered feedback on interface and functionality to assess user acceptance. Palau \cite{MFP2024} conducted similar tests for its stablecoin system. The Bahamas' “Sand Dollar” project \cite{central_bank_of_the_bahamas_2024} highlights CBDC's role in promoting financial inclusion and stresses the need for robust regulatory frameworks to ensure financial stability and security.

\begin{table}[ht!]
    \tiny
    \centering
    \begin{tabularx}{1\textwidth}{|p{1.3cm}|p{1.4cm}|p{0.632cm}|p{1.9cm}|p{1.4cm}|p{2cm}|p{1.8cm}|}
    \hline
    \rowcolor{gray!10}
    \textbf{CBDC Projects} &  \textbf{Country/ Area} &  \textbf{Status} &  \textbf{Architecture} &  \textbf{Ledger Technology} &  \textbf{Access  Model} &  \textbf{Application  Area} \\ \hline
        Sand Dollar & Bahamas & Issued & 2-tier \cite{auer2022central, genc2024literature, noll2024observations} & DLT \cite{auer2022central, noll2024observations}& Account-based \cite{noll2024observations}& Retail \cite{noll2024observations} \\ \hline
        ENaira & Nigeria & Issued & 2-tier \cite{ahiabenu2022comparative} & DLT \cite{ahiabenu2022comparative}& Token-based \cite{ahiabenu2022comparative} & Retail \cite{ahiabenu2022comparative, genc2024literature}  \\ \hline
        Jam-Dex & Jamaica & Issued & 2-tier \cite{noll2024observations} & CT \cite{noll2024observations} & Token-based \cite{noll2024observations} & Retail \cite{noll2024observations}  \\ \hline
        DCash & Eastern Caribbean & Pilot & 2-tier \cite{boar2020impending, noll2024observations} & DLT \cite{boar2020impending, noll2024observations, shapoval2020central} & Token-based \cite{boar2020impending, noll2024observations, shapoval2020central} & Retail \cite{boar2020impending, noll2024observations, shapoval2020central} \\ \hline
        E-CNY & China & Pilot & 2-tier \cite{auer2022central, auer2020rise} & Hybrid \cite{auer2020rise} & Hybrid \cite{auer2020rise} & Hybrid \cite{kalal2023comparative, auer2020rise} \\ \hline
        E-Cedi & Ghana & Pilot & 2-tier \cite{ahiabenu2022comparative, GieseckeDevrient2023, BankOfGhana2023} & CT \cite{GieseckeDevrient2023} & Token-based \cite{ahiabenu2022comparative, BankOfGhana2023} & Retail \cite{ahiabenu2022comparative, BankOfGhana2023}  \\ \hline
        Digital  Tenge & Kazakhstan & Pilot & 2-tier \cite{DigitalTenge2023} & DLT \cite{DigitalTenge2023} & Token-based \cite{DigitalTenge2023} & Hybrid \cite{DigitalTenge2023}  \\ \hline
        Digital Rupee & India & Pilot & 2-tier \cite{dumbre2024blockchain, IndiaCBDC2022} & DLT \cite{dumbre2024blockchain} & Hybrid \cite{dumbre2024blockchain, IndiaCBDC2022} & Hybrid \cite{IndiaCBDC2022}  \\ \hline
        \multirow{2}{*}{Project Aber} & United Arab Emirates & \multirow{2}{*}{Pilot} & \multirow{2}{*}{1-tire  \cite{Aber2020}} & \multirow{2}{*}{DLT \cite{Aber2020}} & \multirow{2}{*}{Account-based \cite{Aber2020}} & \multirow{2}{*}{Wholesale \cite{Aber2020}}  \\ \cline{2-2} 
        ~& Saudi Arabia & ~ &~& ~ & ~ &~ \\\hline
        Helvetia& Switzerland & Pilot& 2-tier \cite{Jordan2024HelvetiaIII} &DLT \cite{guo2024dlt, Jordan2024HelvetiaIII}&Token-based \cite{guo2024dlt, Jordan2024HelvetiaIII} &Wholesale \cite{guo2024dlt, Jordan2024HelvetiaIII} \\ \hline
        Project Khokha & South Africa & Pilot & 2-tier \cite{ProjectKhokha2} & DLT \cite{ProjectKhokha2} & Hybrid \cite{ProjectKhokha2} & Wholesale \cite{ProjectKhokha2}  \\ \hline
        Student Safe & Hungary & Pilot & 1-tire \cite{Faykiss2024} & CT \cite{Faykiss2024} & Account-based \cite{Faykiss2024} & Retail \cite{Faykiss2024}  \\ \hline
        \multirow{5}{*}{Mbridge} & China & \multirow{5}{*}{Pilot} & \multirow{5}{*}{2-tier \cite{mBridge2023}} & \multirow{5}{*}{DLT \cite{mBridge2023}}  & \multirow{5}{*}{Hybrid \cite{mBridge2023}}  & \multirow{5}{*}{Wholesale \cite{mBridge2023}} \\ \cline{2-2}  & Hongkong & & & & & \\ \cline{2-2}  & Saudi Arabia & & & & &  \\ \cline{2-2}  & Thailand  & & & & & \\ \cline{2-2}  & United Arab Emirates & & & & & \\\hline
        Digital Yen & Japan & Pilot & 2-tier \cite{Bank_of_Japan_2024} & Hybrid \cite{Bank_of_Japan_2024} & Account-based \cite{Bank_of_Japan_2024} & Retail \cite{Bank_of_Japan_2024}  \\ \hline
        Digital Lilangeni & Eswatini & Pilot & 2-tier \cite{CBE2024}& Hybrid \cite{CBE2024} & Token-based \cite{CBE2024} & Retail \cite{CBE2024} \\ \hline
        E-AUD & Australia & Pilot & 2-tier \cite{RBA2024} & DLT \cite{RBA2024} & Token-based \cite{cbdctracker} & Hybrid \cite{RBA2024} \\ \hline
        Palau Stablecoin & Palau & Pilot & 2-tier \cite{MFP2024}& DLT \cite{MFP2024} & Token-based \cite{MFP2024} & Hybrid \cite{atlanticcouncil_cbdctracker}  \\ \hline
        Digital Baht & Thailand  & Pilot & 2-tier \cite{BOT_2024} & Hybrid \cite{BOT_2024} & Token-based \cite{BOT_2024}  & Hybrid \cite{BOT_2024}  \\ \hline
         E-HKD & Hong Kong,China & Pilot & 2-tier \cite{HKMA_2021} & Hybrid \cite{HKMA_2021} & Hybrid \cite{HKMA_2021} & Hybrid \cite{atlanticcouncil_cbdctracker}  \\ \hline
        Digital Won & South Korea & Pilot & 2-tier \cite{atlanticcouncil_cbdctracker}& DLT \cite{atlanticcouncil_cbdctracker} & Token-based \cite{atlanticcouncil_cbdctracker} & Hybrid \cite{atlanticcouncil_cbdctracker}  \\ \hline
        Digital ruble & Russia & Pilot & 2-tier \cite{Smirnova_2023} & Hybrid \cite{Smirnova_2023} & Account-based \cite{Smirnova_2023} & Hybrid \cite{Smirnova_2023} \\ \hline
        Digital Lira & Turkey & Pilot & 2-tier \cite{CBRT2023_digitalturkishlira} & DLT \cite{CBRT2023_digitalturkishlira} & Account-based \cite{CBRT2023_digitalturkishlira} & Retail \cite{cbdctracker}  \\ \hline
        E-hryvnia & Ukraine & Pilot & 2-tier \cite{NBU2021_ehryvnia} & DLT \cite{NBU2018_ehryvnia} & Hybrid \cite{NBU2018_ehryvnia} & Retail \cite{NBU2021_ehryvnia}  \\ \hline
        E-krona & Sweden & Pilot & 2-tier \cite{Riksbank2023_ekronapilot} & DLT \cite{Riksbank2023_ekronapilot} & Token-based \cite{Riksbank2023_ekronapilot} & Retail \cite{Riksbank2023_ekronapilot} \\ \hline
        Digital Euro & Spain & Pilot & 2-tier \cite{atlanticcouncil_cbdctracker} & DLT \cite{atlanticcouncil_cbdctracker} & Token-based \cite{atlanticcouncil_cbdctracker} & Wholesale \cite{atlanticcouncil_cbdctracker}  \\ \hline
        DREX & Brazil & Pilot & 2-tier \cite{drex_forum_2023} & DLT \cite{drex_forum_2023} & Token-based \cite{drex_forum_2023} & hybrid \cite{BancoCentraldoBrasil_2025}  \\ \hline
    \end{tabularx}
    \caption{CBDC in Different Countries/Areas.}
    \label{CBDC in the world}
\end{table}
\subsection{Issued and Pilot CBDC}

Ultimately, 26 complete design schemes were identified across regions and organizations. Based on the outlined design options, Table \ref{CBDC in the world} surveys CBDC projects by system architecture, ledger technology, access model, and application areas. Of these programs, three of them have already been issued and the remaining 23 are being piloted.\\

\begin{figure} 
    \centering
    \centerline{\includegraphics[height=7cm]{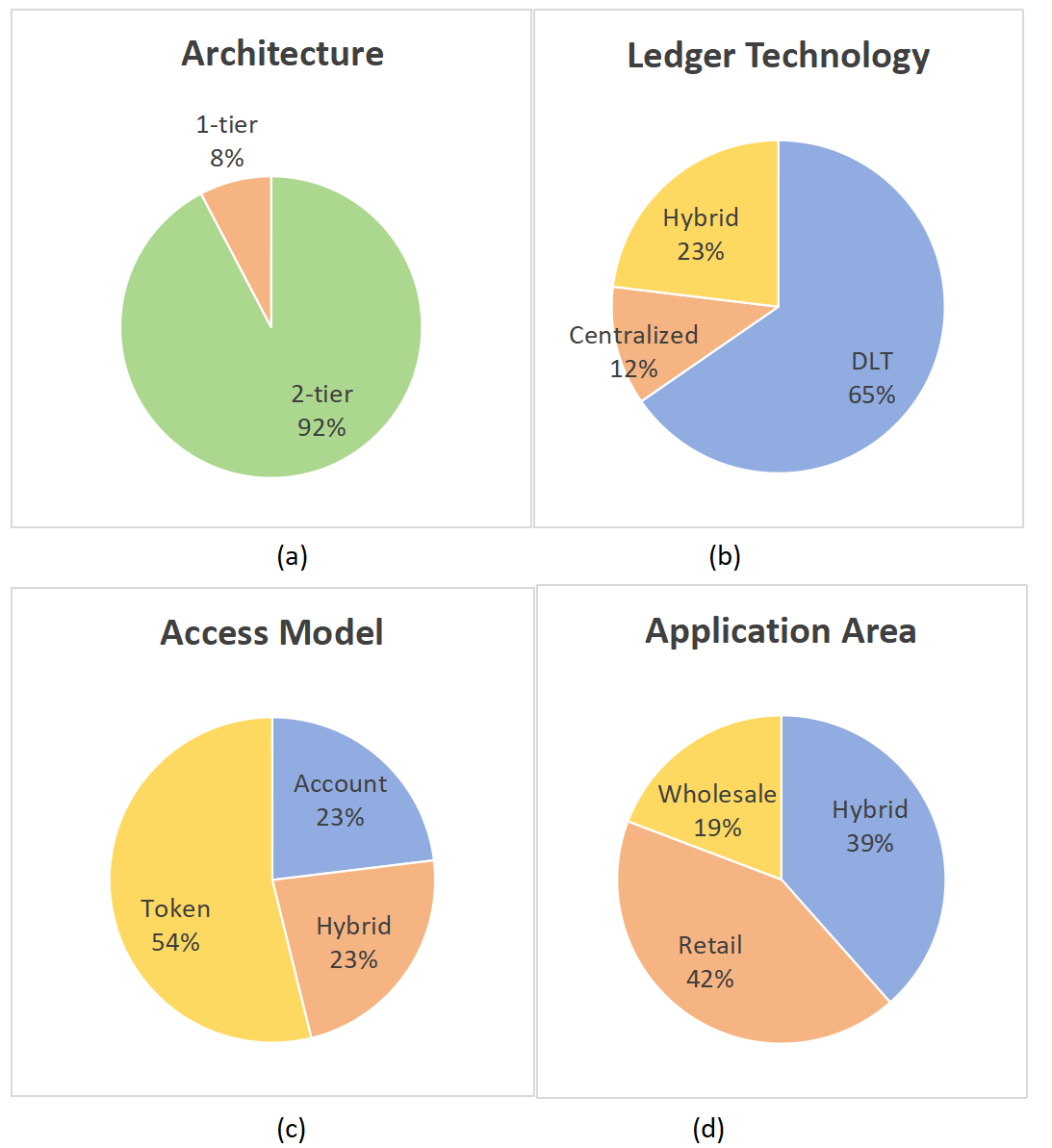} } 
    \caption{Overview of Each Design Option} 
    \label{fig 6} 
\end{figure}

Based on the 26 projects in Table \ref{CBDC in the world}, their main design options are summarized in Figure \ref{fig 6}.The results show that 92\% of countries adopted a 2-tier architecture, while only 8\% chose a 1-tier model, as shown in Figure \ref{fig 6}(a). All three countries that have issued CBDCs use the 2-tier model with intermediary access. As shown in Figure \ref{fig 6}(b), 65\% of projects adopted DLT, 23\% used a hybrid of centralized and distributed technologies, and only three countries chose a traditional centralized ledger, including the issued Jam-Dex. Similarly, in Figure \ref{fig 6}(c), over half the projects adopted a token-based access model, while 23\% used account-based models, and another 23\% implemented a hybrid of both. Regarding application areas, Figure \ref{fig 6}(d) shows most CBDC projects target retail (42\%), followed by combined retail and wholesale (39\%), and exclusively wholesale applications (19\%). The above analysis indicates that 2-tier, DLT-based, and token-based designs dominate both implemented and pilot CBDC projects worldwide.\\

It is worth noting that Helvetia III, the third phase of Switzerland's Helvetia project (Table \ref{CBDC in the world}), represents a pioneering pilot of this model and marks the world's first issuance of a wholesale CBDC (wCBDC) on a regulated third-party platform. On the SIX Digital Exchange (SDX), the project issued a single-currency wCBDC for settling tokenized asset transactions within Switzerland. It successfully implemented multiple tokenized bond issuances and secondary market transactions, while addressing challenges such as governance, currency fragmentation, and future connectivity with other payment systems, alongside the integration of privacy-preserving technologies.



The project adopts a two-tier architecture: the central bank issues the CBDC, while private-sector platforms handle its distribution and transaction processing. It is built on Corda DLT, designed for financial institutions and regulated environments, using a permissioned ledger accessible only to entities in Switzerland's RTGS system. To prevent double spending, a notary service validates transaction inputs and updates the ledger state accordingly. Smart contract functionality supports complex transaction logic and automates processes such as issuance, trading, and settlement of tokenized assets.



Beyond technological advancement, Helvetia III ensures seamless integration with the Swiss RTGS, enabling smooth asset conversion and transaction processing from traditional systems to distributed ledgers. This design enhances currency stability and security while minimizing reconciliation requirements. To mitigate the risk of fragmentation — especially when multiple third-party platforms are involved — the project employs standardized automation and aligns SDX operating hours with the RTGS system.


\subsection{Challenges in CBDC Implementation}
Although many countries have issued or piloted CBDCs, overall adoption remains low. As of 2023, Jam-Dex accounted for just 0.1\% of money in circulation, and e-yuan represented only 0.16\% of the M0 supply. By August 2024, Sand Dollar circulation reached \$2.39 million—just 0.5\% of banknotes—with per capita usage below \$3 \cite{central_bank_of_the_bahamas_2024}. Several factors contribute to this limited uptake.


A key concern with CBDCs is their potential vulnerability to cyber-attacks due to underlying technologies \cite{CORBET2024106226}. Design-related technical challenges are real and require urgent attention. Australia's retail CBDC study \cite{RBA2024} noted that while cryptography enhances security, it also increases computational load, reducing transaction speed. Issuing larger token denominations may alleviate this but does not resolve the core efficiency bottleneck.


Additionally, the choice of ledger technology, particularly whether to adopt DLT, requires careful evaluation, as it may reduce transaction speed and limit scalability. Additionally, enabling fully offline payments and managing risks from programmability are critical challenges that must be addressed in CBDC design.


While research on CBDCs is expanding, several projects have stalled or been discontinued. As early as 2014, Finland implemented a CBDC-like system, but reliance on costly smart cards led to its termination. In 2018, El Salvador abandoned its retail CBDC due to low demand and trust issues \cite{bespalova2024crypto}. In 2024, both Canada and Australia suspended their CBDC plans \cite{bakx2024digital}; a Bank of Canada survey showed 90\% of respondents were concerned about privacy and fund security, and 92\% were unwilling to use a digital Canadian dollar \cite{digital2023canadian}. Similarly, the U.S. banned CBDC issuance and use in January 2025 \cite{whitehouse2025presidential}.


Unsurprisingly, most of these projects are retail-focused, in countries where alternative digital payment solutions are already widely available. Existing systems often meet daily transaction needs, while users remain concerned about CBDC technical design and privacy security.


\section{Discussion} 

\subsection{Privacy Protection in CBDC System} 
Numerous studies have highlighted that privacy is a critical consideration in the design of CBDC systems \cite{choi2023central, grothoff2021issue, freiman2024cbdc, li2025current}. Unlike cash, CBDCs are based on a digital ledger, meaning that transaction records can be tracked and monitored \cite{brunnermeierdigital}.  Common methods include, but are not limited to, the following:

\begin{itemize}
    \item 
    Privacy-Enhancing Technologies (PETs), including Zero-Knowledge Proofs, Homomorphic Encryption, and Differential Privacy, support secure transactions without disclosing user information. These techniques are typically embedded at the architectural level, though they may be error-prone and require rigorous design\cite{freiman2024cbdc, 10.1145/3561796}.

    \item To protect privacy, CBDC systems should separate personally identifiable information from transaction data and minimize the visibility of user data to central authorities \cite{narula134digital, brunnermeierdigital}. Centralized storage heightens the risk of misuse, making data segmentation a key strategy for reducing attack surfaces.


    \item CBDC design must balance legal compliance with user anonymity. Government access to personal data should be limited to lawful and clearly defined circumstances to prevent excessive surveillance\cite{freiman2024cbdc}.
    


    \item Enhance the right of users to control their data. CBDCs should be designed to empower users by ensuring they have control over how their data is accessed, shared, and transferred \cite{freiman2024cbdc}.
    
\end{itemize}

The programmability of a CBDC can influence privacy protection by enabling the enforcement of predefined rules and conditions, potentially allowing for the monitoring or restriction of user behavior \cite{freiman2024cbdc}. While this functionality may support policy implementation, it must be carefully designed to safeguard user privacy and comply with legal and regulatory requirements.


CBDC privacy protection must address both data confidentiality and resilience against security threats \cite{freiman2024cbdc}, including tampering, loss, and unauthorized disclosure. Although DLT can enhance security, many public blockchains (e.g., Bitcoin, Ethereum) expose sender and receiver addresses \cite{10.1145/3548606.3560707}, making users vulnerable to re-identification via network analysis, address clustering, and transaction fingerprinting \cite{feng2019survey, lee2021survey}.

In the design and implementation of CBDC systems, balancing personal data protection, transaction traceability, and regulatory compliance remains a central challenge. To address this, innovative liquidity-saving mechanisms should be continuously explored to reconcile privacy with auditability—particularly in real-time settlement environments, where post hoc transaction tracing becomes increasingly difficult\cite{li2025current}.


\subsection{Social, Institutional, and Monetary Implications}
\subsubsection{Impacts on Payment Systems and Financial Inclusion} 

Central Bank Digital Currencies (CBDCs) are poised to transform payment systems by integrating advanced technologies such as Distributed Ledger Technology (DLT) and blockchain. Originally deployed in private-sector applications, these technologies are now increasingly adopted in sovereign digital currency designs to support secure, transparent, and interoperable transaction infrastructures. Notably, cross-border CBDC initiatives—such as Project mBridge—are accelerating innovation in DLT by focusing on interoperability and multi-jurisdictional coordination \cite{Horvath2021Concept}.

One of the most significant areas of impact is cross-border payment efficiency. Traditional systems typically rely on chains of correspondent banks, each handling reconciliation, settlement, and compliance, leading to high costs, delays, and operational complexity. In contrast, wholesale CBDCs can integrate multiple national currencies into a unified settlement layer, minimizing the role of intermediaries and enabling near-instantaneous, low-cost cross-border transfers\cite{auer2022central}. By simplifying architecture and enhancing traceability, CBDCs offer a promising alternative to legacy systems.

In addition to improving systemic efficiency, CBDCs also offer opportunities to address long-standing issues in financial inclusion. The adoption of CBDC systems is expected to significantly promote financial inclusion in some countries \cite{BIS2021AER, auer2022central}. Many individuals lack access to banking services due to high account-opening barriers or geographic constraints \cite{BIS2021AER}. With lower adoption costs, CBDCs can reach underserved populations by simplifying authentication and enabling community-based access. However, the persistence of the digital divide highlights the need to ensure that vulnerable groups—such as the elderly and rural populations—can effectively access these systems. This requires careful attention to device compatibility, user interface design, and offline accessibility\cite{li2025current}.

\subsubsection{Implications for the Banking Sector and Monetary Policy} 
Beyond their impact on payment systems and inclusion, CBDCs also pose structural implications for the traditional banking sector. As central bank liabilities, CBDCs offer high security and could draw deposits away from commercial banks \cite{auer2022central}. To retain customers, banks may raise deposit rates, increasing operational costs. In times of economic instability, mass conversion of deposits to CBDCs could heighten liquidity and bankruptcy risks \cite{BIS2021AER}. Conversely, CBDCs create new opportunities for Non-Bank Financial Institutions (NBFIs) to access the system and expand service offerings \cite{auer2022central}.


By reducing reliance on commercial intermediaries, CBDCs may facilitate a more direct and efficient transmission of monetary policy \cite{auer2022central}. They may also help overcome the zero interest rate floor—the constraint that nominal rates cannot fall below zero without prompting cash hoarding. During recessions or deflation, this limits central banks' ability to stimulate demand. CBDCs enable the design of incentives that reduce reliance on cash, offering greater flexibility and enhancing policy effectiveness.


\section{Conclusion} 
This paper surveys research conducted between 2016 and 2025 on the design, implementation, and impact of CBDC systems. It analyzes the multidimensional design of CBDC systems across four main dimensions: architecture, ledger technology, access model, and application areas. Additionally, the paper summarizes and analyzes current design options for CBDC systems, concluding that the most popular CBDC system design, based on current research, is a combination of a two-tier architecture, token-based access, and distributed ledger technology. The paper also emphasizes the critical balance between privacy protection and legal compliance. However, most current CBDC systems are still in experimental pilot phases rather than large-scale deployments, and actual application data may differ in the future. Furthermore, the application of CBDC in cross-border payments may become a primary focus for central banks. If CBDC is to be adopted globally, it must be driven by multiple benefits while also balancing its relationship with existing financial institutions and private payment systems. Concerns regarding the technology behind CBDC systems remain a major barrier to their widespread adoption. Future research should focus on addressing these challenges. This survey provides a theoretical foundation for the design and implementation of CBDC systems, serving as a catalyst for optimizing their design and advancing CBDC development.

\section*{Acknowledgments}
This research was funded by the University of Macau (file no. MYRG2022-00162-FST and MYRG2019-00136-FST).

\bibliographystyle{elsarticle-num} 
\bibliography{ref.bib}
\end{document}